\documentclass[aps,preprint]{revtex4}

\usepackage{graphics,graphicx}
\usepackage{amsmath,amssymb}
\usepackage{epsfig}
\usepackage{txfonts}
\usepackage{subfigure}
\usepackage{xcolor}
\usepackage{epstopdf}
\usepackage{hyperref}

\begin{document}

\title{Scalar particle production in a simple Horndeski theory}

\author{Hao Yu$^a$\footnote{yuh13@lzu.edu.cn},
        Wen-Di Guo$^a$\footnote{guowd14@lzu.edu.cn},
        Ke Yang$^{b}$\footnote{keyang@swu.edu.cn},
        and Yu-Xiao Liu$^a$\footnote{liuyx@lzu.edu.cn, corresponding author}}

\affiliation{
$^a$Institute of Theoretical Physics, Lanzhou University, Lanzhou 730000, People's Republic of China\\
$^b$School of Physical Science and Technology, Southwest University, Chongqing 400715, People's Republic of China}

\begin{abstract}
{The scalar particle production through a scalar field non-minimally coupled with geometry is investigated in the context of a spatially homogeneous and isotropic universe. In this paper, in order to study the evolution of particle production over time in the case of analytical solutions, we focus on a simple Horndeski theory. We first suppose that the universe is dominated by a scalar field and {derive the energy conservation condition}. Then from the thermodynamic point of view, the macroscopic non-conservation of the scalar field energy-momentum tensor can be explained as an irreversible production of the scalar particles. Based on the explanation, {we obtain} a scalar particle production rate and the corresponding entropy. Finally, since the universe, in general, could be regarded as a closed system satisfying the laws of thermodynamics, {we naturally} impose some thermodynamic constraints {on it}. {The} thermodynamic properties of the universe can provide additional constraints on the simple Horndeski theory.
}
\end{abstract}



\maketitle

\linespread{1.2}

\section{Introduction}

Exact tests of the equivalence principle have never ceased~\cite{Damour:1996xt,Gasperini:1988zf,Bertolami:2007zm,Schlamminger:2007ht,Damour:2002nv}
(refer to Ref.~\cite{Will:2014kxa} for the latest review).
Since there have always been researches and observations claiming that violation of the equivalence principle may be really reasonable~\cite{Halprin:1995vg,Adunas:2000zn,Damour:2002nv,Bertolami:2007zm}, the modified gravity theories with violation of the equivalence principle are not only alive but also have attracted increasing attention. For example, in recent years, a kind of modified gravity theory taking into account non-minimal couplings between geometry and matter has become one of the mainstream modified gravity theories (a detailed introduction can be found in Ref.~\cite{Harko:2014gwa}).

Early applications of the non-minimal couplings of matter with gravity were mainly to solve the cosmological constant problem~\cite{Dolgov:2003fw,Mukohyama:2003nw} and accelerated expansion of the universe~\cite{Nojiri:2004bi,Allemandi:2005qs}. Later, Bertolami and coworkers pointed out that if there is a non-minimal coupling of matter to a function of curvature, the trajectory of a test particle in gravitational field will be non-geodesic. The effect of the non-minimal coupling is equivalent to an extra force on the test particle~\cite{Bertolami:2007gv}. In the same year, Faraoni studied the stability condition for the gravity theory~\cite{Faraoni:2007sn}. Since then, related researches mainly focus on galactic rotation curves~\cite{Sotiriou:2008it,Bertolami:2009ic,Harko:2010vs}, accelerated expansion~\cite{Bertolami:2010cw,Bisabr:2012tg,Harko:2014aja} and thermodynamics~\cite{Wu:2012ia,Sharif:2013tha,Harko:2014pqa,Momeni:2015fyt}. Recently, this kind of modified gravity theory was injected new vitality because of the researches on particle production~\cite{Harko:2014pqa,Harko:2015pma,Singh:2015oiw} (we will come back to  {this point} later).

The idea that elementary particles may be created during the evolution of the universe can be dated back to the middle of last century. In 1948, Bondi, Gold~\cite{Bondi:1948qk} and Hoyle~\cite{Hoyle:1948zz} suggested, independently, that the universe inevitably possesses a continuous particle production during the process of expansion, which is based on the hypothesis of Dirac~\cite{Dirac:1937ti}. They found that the rate of particle production can hardly be observed directly in their models. In the late 1960s and early 1970s,
Parker did {a lot of work} on the mechanism of particle production from a microscopic point of view~\cite{Parker:1968mv,Parker:1969au,Parker:1971pt}. {He} systematically studied the relation between elementary particles and gravitational field, and pointed out that the production of elementary particles mainly occurred at the early stage of the universe. During the same period, Tyron~\cite{Tryon:1973xi} and  Fomin~\cite{Fomin:1973xi} also {independently} proposed an interesting approach to create universe or particles, in which our universe comes from a fluctuation of the vacuum. After that, most studies on particle production were related to the applications in cosmology~\cite{Sexl:1969ix,Zeldovich:1970si,Fulling:1974pu,Grib:1976pw,Brout:1977ix,
Hu:1978zd,Papastamatiou:1979rv}.

By the mid-1980s, Prigogine et al., for the first time, considered the correlation between matter and entropy by incorporating the entropy of particle production into Einstein's field equations~\cite{Prigogine:1986zz,Prigogine:1988zz,Prigogine:1989zz}. Applying the result in conjunction with the second law of thermodynamics to cosmology, they found it could address the singularity problem of the universe through an irreversible particle production. Soon, Calv$\tilde{a}$o and collaborators indicated that as long as the specific entropy is not a constant, particles can also be annihilated while the universe satisfies the entropy increase principle~\cite{Calvao:1991wg}. Since {particle production} behaves as a negative pressure term in Einstein's field equations, some researchers also used particle production to  generate inflation~\cite{Ford:1986sy,Traschen:1990sw,Abramo:1996ip,Lima:1995xz,Gunzig:1997tk,Peebles:1998qn,
Zimdahl:1999tn} and {to explain} current cosmic acceleration~\cite{Steigman:2008bc,Lima:2009ic,Allahverdi:2010xz,Jesus:2011ek,Chakraborty:2014fia,
Nunes:2015rea}.

Recently, particle production based on different mechanisms has entered a new development. In Refs.~\cite{Harko:2014pqa,Harko:2015pma}, the authors considered a non-minimal coupling of matter with functions of curvature in the context of the universe. Combining with the thermodynamic properties of the universe, they defined a particle production rate and the corresponding entropy. Then, {the} particle production rate, production pressure and entropy can be given by abstract expressions with the matter Lagrangian and the functions of curvature. According to the second law of thermodynamics of the isolated system ($dS/dt>0$) and the common sense that a macroscopic isolated system always tends spontaneously to thermodynamic equilibrium ($d^2S/dt^2<0$), they briefly discussed how to use these requirements to restrict particle production.

In this paper, we follow Harko's works and continue to study particle production in the context of the universe. Since it is not necessary to demand that the universe tends to equilibrium state at the early and middle stages, we modify the thermodynamic equilibrium condition $d^2S/dt^2<0$ (for all $t$)~\cite{Callen:xx} to $d^2S/dt^2\rightarrow0$ (when $t\rightarrow\infty$). When the solution of the scale factor does not represent the late stage of the universe, the equilibrium condition can be ignored. We consider a specific matter (scalar field) coupling with geometry. The benefit and motivation are that one can obtain a non-conservative expression of the matter energy-momentum tensor directly from the equation of motion of the matter, and further study could be done in the case of analytical solutions. It also makes up for the shortcoming of previous works, in which it is inconvenient to study the evolution of the entropy at all stages of the universe.

To investigate particle production in the case of analytical solutions, we consider a simple Horndeski theory, which only contains second order derivatives in the equations of motion~\cite{Horndeski:1974wa}. We hope that there will be some new constraints on the coupling coefficients in the simple Horndeski theory by studying the thermodynamic properties of the universe.

The organization of this paper is as follows. In Sec.2, a simple Horndeski theory as our major research object is introduced. In Sec.3, we calculate abstract {expression} of particle production rate from the point of view of thermodynamics. In Sec.4, we solve the equations of motion and constrain the coupling parameters with the laws of thermodynamics. In Sec.5, we briefly discuss the entropy of the apparent horizon. Conclusion and discussion {are} drawn in last section.

\section{A simple Horndeski theory}
\label{MainModel1}

We consider a spatially homogeneous and isotropic Friedmann-Robertson-Walker (FRW) metric, which is given by
\begin{eqnarray}
ds^{2}=-dt^2+a^2(t)\gamma_{ij}dx^i dx^j.\label{metric}
\end{eqnarray}
Here, $a(t)$ denotes a scale factor, which is a function of the cosmic time $t$, and $\gamma_{ij}$ is the metric of a locally homogeneous three-dimensional space with constant curvature $k$.
For simplicity, we only discuss the case of $k=0$ in this paper.

According to Harko's works~\cite{Harko:2014pqa,Harko:2015pma}, particle production could be achieved through a non-minimal coupling between matter and geometry. As we know, the Horndeski theory is a theory of gravity non-minimally coupled to a scalar field {and it} also guarantees that equations of motion are second order~\cite{Horndeski:1974wa}. In order to research particle production in the presence of analytical solutions, we consider a simple Horndeski theory. As a toy model, the action that should be as simple as possible and also representative. In a system of units with $8\pi G=c=1$, the action that we pick out from the whole action of the Horndeski theory has the form
\begin{eqnarray}\label{actionH}
S=\frac{1}{2}\int d^4x \sqrt {-g}\left[R+\lambda_1 G_{\mu\nu}\nabla^\mu\phi\nabla^\nu\phi+\lambda_2F(\phi)R\right]+\int d^4x \sqrt {-g}L_\phi,\label{action3}
\end{eqnarray}
where $G_{\mu\nu}=R_{\mu\nu}-\frac{1}{2}g_{\mu\nu}R$ is the Einstein tensor, $F(\phi)$ is an arbitrary function of $\phi$, $L_\phi=-\frac{1}{2}(\nabla\phi)^2-V(\phi)$ is the Lagrangian density of the scalar field, {and $\lambda_1$ and $\lambda_2$ are real parameters}. For a cosmological model with the FRW metric~(\ref{metric}), the scalar field $\phi$ only depends on the cosmic time, so $L_\phi=\frac{1}{2}\dot{\phi}^2-V(\phi)$. In what follows, a dot over any quantity denotes a derivative with respect to the cosmic time. Note that we neglect other matter fields since they minimally couple to geometry. Though this assumption is not very realistic, it is helpful to simplify the issues that we will discuss and {to get} analytical solutions.

{The energy-momentum tensor of the scalar field is defined as
\begin{eqnarray}
T_{\mu\nu}\equiv-\frac{2}{\sqrt{-g}}\frac{\delta(\sqrt{-g}L_\phi)}{\delta g^{\mu\nu}}=\left(-\frac{1}{2}(\nabla\phi)^2-V(\phi)\right)g_{\mu\nu}+\partial_\mu\phi
\partial_\nu\phi.\label{Tuv}
\end{eqnarray}
Then, the energy density and pressure of the scalar field are given by
\begin{subequations}
\begin{eqnarray}
\label{rho}
\rho_\phi&=&\frac{1}{2}\dot{\phi}^2+V(\phi),\\
\label{p}
p_\phi&=&\frac{1}{2}\dot{\phi}^2-V(\phi).
\end{eqnarray}
\end{subequations}}
The energy conservation law of the scalar field indicates that $T_{\mu\nu}$, in most cases, should satisfy the following relation:
\begin{eqnarray}
\nabla^\mu T_{\mu0}=-\dot{\rho}_{\phi}-3H(p_{\phi}+\rho_{\phi})=
-\dot{\phi}\ddot{\phi}-\dot{V}(\phi)-3H\dot{\phi}^2=0.\label{equtinon25}
\end{eqnarray}

The variations of the action (\ref{action3}) with respect to $g_{\mu\nu}$ and $\phi$ lead to the following field equations
\begin{subequations}
\begin{eqnarray}
&&R_{\mu\nu}-\frac{1}{2}Rg_{\mu\nu}+\lambda_1\bigg\{-\frac{1}{2}R\nabla_\mu\phi\nabla_\nu\phi +2\nabla_\alpha\phi\nabla_{(\mu}\phi R^\alpha_{\nu)}-\frac{1}{2}(\nabla\phi)^2 G_{\mu\nu} -\Box\phi\nabla_\mu\nabla_\nu\phi\nonumber\\
&&+\nabla_\mu\nabla^\alpha\phi\nabla_\nu\nabla_\alpha\phi
+R_{\mu\alpha\nu\beta}\nabla^\alpha\phi\nabla^\beta\phi
+g_{\mu\nu}\Big[\frac{1}{2}(\Box\phi)^2-\frac{1}{2}\nabla^\alpha\nabla^\beta\phi\nabla_\alpha\nabla_\beta\phi
-\nabla_\alpha\phi\nabla_\beta\phi R^{\alpha\beta}\Big]
\bigg\}\nonumber\\
&&+\lambda_2\Big[R_{\mu\nu}F(\phi)-\frac{1}{2}R F(\phi)g_{\mu\nu}+
(g_{\mu\nu}\Box-\nabla_\mu\nabla_\nu)F(\phi)\Big]=T_{\mu\nu},\label{Eqofmotion111}\\
&&(1-3\lambda_1 H^2)\ddot\phi+3H\Big(1-\lambda_1 H^2-2\lambda_1\frac{\ddot a}{a}\Big)\dot\phi+V'(\phi)+6\lambda_2\Big(\frac{\ddot a}{a}+H^2\Big)F'(\phi)=0,\label{horndeskiscalar}
\end{eqnarray}\label{horndeskiscalar11}
\end{subequations}
where $H=\dot{a}(t)/a(t)$ is the Hubble parameter, $V'(\phi)\equiv\frac{dV(\phi)}{d\phi}=\frac{\dot{V}(\phi)}{\dot{\phi}}$, $F'(\phi)\equiv\frac{dF(\phi)}{d\phi}=\frac{\dot{F}(\phi)}{\dot{\phi}}$, and $\Box=g^{\mu\nu}\nabla_\mu\nabla_\nu$.
Taking the relations $\dot{\phi}^2=\rho_{\phi}+p_{\phi}$ and $\dot{\phi}\ddot{\phi}=\frac{1}{2}(\dot{\rho}_{\phi}+\dot{p}_{\phi})$ into Eq.~(\ref{horndeskiscalar}), we have
\begin{eqnarray}
\dot{\rho}_{\phi}+3H(p_{\phi}+\rho_{\phi})&=&\frac{1}{1-\lambda_1 H^2-2\lambda_1 \frac{\ddot a}{a}}\Big[2\lambda_1(H^2-\frac{\ddot a}{a})\ddot\phi\dot\phi-\lambda_1 ( H^2+2\frac{\ddot a}{a} )\dot{V}(\phi) \nonumber\\
&-&6\lambda_2\Big(H^2+\frac{\ddot a}{a}\Big)\dot F(\phi)\Big].\label{hornconv}
\end{eqnarray}
Note that Eq.~(\ref{hornconv}) can also be derived from Eq.~(\ref{Eqofmotion111}). But if we start from Eq.~(\ref{Eqofmotion111}), there will be a lot of trouble. In Refs.~\cite{Harko:2014pqa,Harko:2015pma,Singh:2015oiw}, there is no explicit equation corresponding to Eq. \eqref{horndeskiscalar} since the Lagrangian of the matter is abstract. To obtain the similar equation, one can only proceed from the gravitational field equations.

The energy conservation condition of the scalar field (which guarantees $\nabla^\mu T_{\mu0}=-\dot{\rho}_{\phi}-3H(p_{\phi}+\rho_{\phi})=0$) is given as
\begin{eqnarray}
\lambda_1(2 H^2-2 \frac{\ddot a}{a})\ddot\phi\dot\phi=\lambda_1( H^2+2 \frac{\ddot a}{a} )\dot{V}(\phi)+6\lambda_2\Big(H^2+\frac{\ddot a}{a}\Big)\dot F(\phi).
\end{eqnarray}
When $\lambda_1=0$, the energy conservation condition is $6\lambda_2\Big(H^2+\frac{\ddot a}{a}\Big)\dot F(\phi)=0$. If $\dot F(\phi)=0$, there is only minimal coupling between the scalar field and geometry. In this case, the energy-momentum tensor of the scalar field naturally satisfies Eq.~(\ref{equtinon25}). If $\dot F(\phi)\neq0$, we need $H^2+\frac{\ddot a}{a}=0$, {for which} the solution of the scale factor is $a(t)= c_1 \sqrt{2 t-c_2}$, where $c_1$ and $c_2$ are constants. Similarly, when $\lambda_2=0$, to guarantee energy conservation of the scalar field, we need $(2 H^2-2 \frac{\ddot a}{a})\ddot\phi\dot\phi=( H^2+2 \frac{\ddot a}{a} )\dot{V}(\phi)$.

It is worth mentioning that since we have defined {the scalar} field energy-momentum tensor by Eq.~(\ref{Tuv}), in general, if there exists {a non-minimal} coupling between the scalar field and geometry, there must be energy exchange between the scalar field and ``space-time" (energy exchange is certainly not equal to zero). But when the energy conservation condition is satisfied, apparently there is no energy flowing into or out of the scalar field. How do we explain this seemingly contradictory description? We can assume that as long as there exists {a non-minimal} coupling between the scalar filed and geometry, all energy of the scalar field (including kinetic energy and potential energy) can exchange with ``space-time". In general, the overall exchange of energy is not equal to zero. However, when the evolutions of the scalar field and background space-time satisfy a particular condition, i.e., the energy conservation condition, the total amount of energy exchange exactly vanishes. The ``space-time" can be considered as a medium, whose role is to achieve internal energy exchange between the potential energy and kinetic energy of the scalar field.

\section{Particle production from the point of view of thermodynamics}
\label{application}
In this section, we study thermodynamic properties of an ``open universe" in {a simple} Horndeski theory. Here, the ``open universe" that we study is actually a closed universe according to the traditional sense of thermodynamics and we could assume that the boundary of the closed universe is the apparent horizon. The word ``open" just means that the objects in an isolated system can exchange energy with the background space-time. Now we start from the point of view of thermodynamics and then define the scalar particle production rate.

For a homogeneous and isotropic universe which contains $N$ particles in a volume $V=a^3$ (we suppose the universe is a perfect fluid), the second law of thermodynamics requires~\cite{Prigogine:1988zz}
\begin{eqnarray}
\frac{d}{dt}(\rho a^3)+p\frac{d}{dt}a^3=\frac{dQ}{dt}+\frac{p+\rho}{n}\frac{d}{dt}(n a^3).\label{secondlaw}
\end{eqnarray}
Here, $n=N/V$ is the particle number density and $dQ/dt$ is the rate of heat transfer into or out of the universe. The parameter $p$ is the sum of pressure of all particles in the universe and $\rho$ is the corresponding energy density. Due to the cosmological principle, the universe is normally regarded as an adiabatic system, i.e, $dQ/dt=0$. Therefore, in the following we ignore heat transfer during the evolution of the universe. Note that ``adiabatic system" is not incompatible with ``open universe''. Here again, the ``open" means the interaction between matter and background space-time, which leads to non-conservation of the matter but the whole energy of the universe should be conserved.

For further simplification, we assume that in such an adiabatic universe there exists a class of scalar {particles}, for example quintessence~\cite{Caldwell:1997ii,Peebles:2002gy,Linder:2007wa}, which almost has no interaction with other particles. We focus on this kind of scalar {particles} and ignore other {ones}. In addition, the bulk viscous pressure and viscous dissipation of the energy-momentum tensor of the scalar field are also neglected in our model. So, in a comoving reference frame the scalar field energy-momentum tensor can be given as $T_{\mu\nu}=(p_\phi+\rho_\phi)U_\mu U_\nu+p_\phi g_{\mu\nu}$, where $U_\mu$ is the four-velocity of a comoving observer satisfying $U_\mu U^\mu=-1$ and $\nabla_\nu U^\mu U_\mu=0$. We reformulate Eq.~(\ref{secondlaw}) in an equivalent form
\begin{eqnarray}
\dot\rho_\phi+3H(\rho_\phi+p_\phi)=\frac{p_\phi+\rho_\phi}{n}(\dot n+3Hn).\label{secondlaw2}
\end{eqnarray}
When the total number of salar particles is a constant, i.e., $\dot n+3Hn=0$, energy conservation holds for the scalar field. If $\dot n+3Hn\neq0$, from the point of view of thermodynamics the total number of the scalar particles is not conserved. {In other} words, there exists scalar particle production or annihilation. Comparing this equation with Eq.~(\ref{hornconv}), in the presence of coupling between geometry and scalar field, the non-conservation of energy of the scalar field can be explained as scalar particle production or annihilation.

We define a scalar particle production rate $\Gamma_\phi$ ($\Gamma_\phi<0$ {means annihilation}) as~\cite{Harko:2014pqa,Harko:2015pma}
\begin{eqnarray}
\dot n+3Hn=\Gamma_\phi n.\label{productionrate0}
\end{eqnarray}
For the simple Horndeski theory, the scalar particle production rate is given by
\begin{eqnarray}\label{gammaH}
\Gamma_\phi=\frac{1}{1\!-\!\lambda_1 H^2\!-\!2 \frac{\lambda_1\ddot a}{a}}\bigg[2\lambda_1 (H^2\!-\!\frac{\ddot a}{a})\frac{\ddot\phi}{\dot\phi}\!-\!\lambda_1\Big(H^2\!+\!\frac{2\ddot a}{a}\Big)\frac{\dot{V}(\phi)}{\dot\phi^2}\!-\!6\lambda_2\Big(H^2\!+\!\frac{\ddot a}{a}\Big)\frac{\dot F(\phi)}{\dot\phi^2}\bigg].\label{productionrate3}
\end{eqnarray}
Giving any solution of Eq.~(\ref{horndeskiscalar11}), we can get the evolution of $\Gamma_\phi$ over the cosmic time.

Then, we need to calculate the entropy of the scalar particles. We ignore the entropy of the spcae-time and the apparent horizon. {So the} total entropy of the universe is just relevant to the scalar particles. The differential expression of the total entropy consists of two parts:
\begin{eqnarray}
dS_{in}=dS_{f}+dS_{c},
\end{eqnarray}
where $dS_{f}$ represents entropy flow and $dS_{c}$ is entropy production. For a stable thermodynamic system without particle production, if it has no energy exchange with outside, we have $dS_{f}=0$, $dS_{c}=0$ and, therefore, $dS_{in}=0$. For the {system we study}, since it could be regarded as an isolated system, only $dS_{f}$ vanishes. So the total entropy can be expressed as~\cite{Prigogine:1988zz,Prigogine:1989zz,Harko:2014pqa,Harko:2015pma} (we suppose {that} the specific entropy of the scalar particle is {a constant})
\begin{eqnarray}\label{ds1}
\frac{dS_{in}}{dt}=\frac{S_{in}}{n_{\phi}}(\dot n_{\phi} +3 H n_{\phi})=\Gamma_\phi S_{in},
\end{eqnarray}
where $n_{\phi}$ is the particle number density of the scalar {particles}. With any given $\Gamma_\phi$, $S_{in}$ can be expressed as a function of {the} cosmic time.

\section{Thermodynamic constraints on the simple Horndeski theory}
In this section, we need to solve the field equations analytically and re-express the scalar particle production rate and the corresponding entropy as functions of the cosmic time. Utilizing the constraints from the laws of thermodynamics we get new constraints on the parameters of the simple Horndeski theory.

Let us retrospect Eq.~(\ref{horndeskiscalar11}), which can be rewritten as
\begin{subequations}
\begin{eqnarray}
\frac{1}{2}\dot\phi^2+V(\phi)&=&-3H^2+\lambda_1\bigg\{ \frac{3}{2}H^2\dot\phi^2-\ddot\phi^2+\ddot\phi\Big(3H\dot\phi+\ddot\phi\Big)-
\Big[\frac{1}{2}\Big(3H\dot\phi+\ddot\phi\Big)^2-\frac{3\ddot a}{a}\dot\phi^2\nonumber\\
&-&\frac{1}{2}\Big(3H^2\dot\phi^2+\ddot\phi^2\Big)
\Big]\bigg\}+\lambda_2\Big[3\frac{\ddot a}{a}F(\phi)-3\Big(H^2+\frac{\ddot a}{a}\Big)F(\phi)+3H\dot F(\phi)\Big],\\
\frac{1}{2}\dot\phi^2-V(\phi)&=&H^2+2\frac{\ddot a}{a}+\lambda_1\bigg\{ \frac{3}{2}H^2\dot\phi^2+2\frac{\ddot a}{a}\dot\phi^2-H\dot\phi\Big(3H\dot\phi+\ddot\phi\Big)+
\Big[\frac{1}{2}\Big(3H\dot\phi+\ddot\phi\Big)^2-\frac{3\ddot a}{a}\dot\phi^2\nonumber\\
&-&\frac{1}{2}\Big(3H^2\dot\phi^2+\ddot\phi^2\Big)\Big]\bigg\}-\lambda_2\Big[\Big(2H^2+\frac{\ddot a}{a}\Big)F(\phi)-3\Big(H^2+\frac{\ddot a}{a}\Big)F(\phi)+\ddot F(\phi)\nonumber\\
&+&2H \dot F(\phi)\Big],\\
V'(\phi)&=&-(1-3\lambda_1 H^2)\ddot\phi-3H\Big(1-\lambda_1 H^2-2\lambda_1\frac{\ddot a}{a}\Big)\dot\phi-6\lambda_2\Big(\frac{\ddot a}{a}+H^2\Big)F'(\phi).
\end{eqnarray}\label{motion4}
\end{subequations}
Since the universe has different equations of state in different epochs, we prefer to choose three extreme cases (dust, radiation, and vacuum energy) to solve our field equations. However, if the scale factor is given in advance, it {might} be difficult to obtain {solutions} of {other} variables. To make the discussion more abundant if there exist other special analytical solutions, we will also study them.

\subsection{$\lambda_1=0$ and $\lambda_2\neq0$}
\label{A}

\subsubsection{$a(t)\propto e^{h t}$}

Let us first study the case $\lambda_1=0$ and $\lambda_2\neq0$. For the stage of inflation or the epoch when the universe is dominated by the vacuum energy, the scale factor grows exponentially with time, i.e., $a(t)\propto e^{h t}$ ($h>0$ is a constant). In this situation we find that the solutions of other variables are trivial:
\begin{subequations}
\begin{eqnarray}
a(t)&\propto& e^{ht},\\
V(\phi)&=&c_1,\\
F(\phi)&=&c_2,\\
\phi(t)&=&c_3,
\end{eqnarray}
\end{subequations}
where $c_i$ ($i=1,2,3$) are constants. { The action (\ref{actionH}) degenerates into the case of minimal coupling gravity and the corresponding scalar particle production rate $\Gamma_{\phi}$ is naturally equal to 0. }

\subsubsection{$a(t)\propto t^{n}$ ($t\neq3/4$)}
For the dust universe and radiation universe, since their scale factors are all polynomials, we can suppose $a(t)\propto t^n$ ($n=2/3$ indicates dust {and $n=1/2$ radiation}) to solve them simultaneously. As {for  other values} of $n$, we discuss some special cases.

We first consider $a(t)\propto t^n$ ($n>0$ and $n\neq3/4$), which results in the following solutions
\begin{subequations}
\begin{eqnarray}
a(t)&\propto& t^{n},\\
V(\phi(t))&=&(3n^2-n)(b_1 \lambda _2+1) \Big(e^{\pm\frac{\sqrt{2} \left(b_2-\phi(t)\right)}{\sqrt{-n \left(b_1 \lambda _2+1\right)}}}\Big)=
-\frac{n(3n-1)(b_1 \lambda_2+1)}{t^2},\\
F(\phi(t))&=&b_1,\\
\phi(t)&=&b_2\pm\sqrt{-2n(b_1 \lambda_2+1)}\log t,
\end{eqnarray}
\end{subequations}
where $b_1$ and $b_2$ are constants and $b_1 \lambda_2+1<0$. Since the solution of $F(\phi)$ is a constant, no matter what value $n$ takes, there is no coupling between the scalar field and geometry, which leads to $\Gamma_{\phi}=0$.

\subsubsection{$a(t)\propto t^{3/4}$}
When we solve the field equations with $a(t)\propto t^{n}$, we find that $a(t)\propto t^{3/4}$ is a special case. {The $V(\phi)$, $\phi(t)$ and $F(\phi)$} just need to satisfy two equations:
\begin{subequations}\label{53}
\begin{eqnarray}
V(\phi(t))&=&\frac{9(-3+\lambda_2(-3F(\phi)+4t\dot F(\phi)))}{16t^2}-\frac12\dot\phi(t)^2,\\
-4t^2\dot\phi(t)^2&=&6+6\lambda_2F(\phi)+\lambda_2 t(4t\ddot F(\phi)-3\dot F(\phi)).
\end{eqnarray}
\end{subequations}
With any given $F(\phi)$, we can obtain $\phi(t)$ and $V(\phi)$ from Eq.~(\ref{53}). If $F(\phi)$ is not {a constant function}, there may exist non-zero particle production rate.  {For the sake of simplicity,} we take $F(\phi)=c_1 t+c_2$ {and then obtain the following solution}
\begin{subequations}\label{eq54}
\begin{eqnarray}
a(t)&\propto& t^{3/4},\\
V(\phi(t))&=&-\frac{15 \big[\lambda _2 (c_2-c_1 t )+1\big]}{16 t^2},\\
F(\phi)&=&c_1 t+c_2,\\
\phi(t)&=&\!\pm\!\sqrt{6 c_2 \lambda _2\!+\!6} \arctan\left(\frac{\sqrt{\!-\!\lambda _2 (2 c_2\!+\!c_1 t )\!-\!2}}{\sqrt{2 c_2 \lambda _2\!+\!2}}\right)\!\mp\!\sqrt{\!-\!3\lambda _2(2 c_2\!+\!c_1 t)\!-\!6},
\end{eqnarray}
\end{subequations}
where $2 c_2 \lambda _2+2>0$ and $c_1\lambda_2 <0$. The range of { the parameter} $t$ is $t>\frac{2 c_2 \lambda _2+2}{- c_1\lambda_2}>0$.
Taking these solutions into the expression of $\Gamma_{\phi}$, we have a non-zero particle production rate:
\begin{eqnarray}\label{gammar}
\Gamma_{\phi}=\frac{3  c_1 \lambda _2}{\lambda _2 (2 c_2+ c_1t)+2}>0.
\end{eqnarray}

Note that for an adiabatic process in a closed thermodynamic system, the irreversible particle production described by Eq.~(\ref{secondlaw}) can be rewritten as an effective conservation equation:
\begin{eqnarray}
\dot\rho_\phi+3H(\rho_\phi+p_\phi+\overline{p}_\phi)=0.
\end{eqnarray}
Here, we define a new thermodynamic quantity $\overline{p}_\phi$, which denotes the production pressure of the scalar particles. Considering solutions (\ref{eq54}), $\overline{p}_\phi$ is given as
\begin{eqnarray}
\overline{p}_\phi=-\frac{\Gamma_{\phi}(p_{\phi}+\rho_{\phi})}{3H}=\frac{c_1 \lambda_2}{t}<0.
\end{eqnarray}
In fact, in some literatures about particle production, the negative production pressure could be used to explain the accelerated expansion of the universe (see early discussions in Refs.~\cite{Prigogine:1989zz,Calvao:1991wg,Lima:1992np}). Through a simple calculation, the total pressure of the scalar particles is given by
\begin{eqnarray}
p_\phi+\overline{p}_\phi=\frac{32t^3 c_1 \lambda_2+30t^2(1+(c_2-c_1t)\lambda_2)-
\frac{3(4+(4c_2+c_1t)\lambda_2)^2}{2+(2c_2+c_1t)\lambda_2}}{32 t^4}.
\end{eqnarray}
Therefore, as long as the parameters are appropriate, the total pressure can be negative and lead to an accelerated expansion.

Now we calculate the total entropy $S_{in}$ (inside of the apparent horizon) of the system and study { its evolution}. Plugging { the particle} production rate~(\ref{gammar}) into Eq.~(\ref{ds1}), the total entropy can be rewritten as
\begin{eqnarray}
\frac{dS_{in}}{S_{in}}= \frac{3  c_1 \lambda _2}{\lambda _2 (2 c_2+ c_1t)+2} dt.
\end{eqnarray}
Then we have
\begin{eqnarray}
S_{in}(t)&=&S_{in}(t_0) \exp\Big[\int_{t_0}^{t} \frac{3  c_1 \lambda _2}{\lambda _2 (2 c_2+ c_1t)+2} dt\Big]\\
&=&S_{in}(t_0) \Big[\frac{\lambda _2 (2 c_2+c_1t)+2}{\lambda _2 (2 c_2+c_1t_0)+2}\Big]^3,\label{entropy}
\end{eqnarray}
where $S_{in}(t_0)$ is the total entropy at time $t_0$.

The second law of thermodynamics states that in a closed thermodynamic system every real process must lead to { increase of the entropy} of the system. In this model with geometry coupling to a scalar field, the space-time background could be considered as a source of energy. From the scalar field point of view, the energy is not conserved. A practical observable effect is the change of the particle number density, which should satisfy {the principle of entropy increase}. In consideration of the second law of thermodynamics, the total entropy $S_{in}(t)$ needs to satisfy
\begin{eqnarray}
\frac{dS_{in}(t)}{dt}=S_{in}(t_0) \frac{3 \lambda _2 c_1 [\lambda _2 (2 c_2+c_1 t )+2]^2}{[\lambda _2 (2 c_2+c_1 t_0 )+2]^3}\geq0.
\end{eqnarray}
Because the total entropy of the system is always non-negative at any time, we have $S_{in}(t_0)>0$. The above constraint condition is equivalent to
\begin{eqnarray}\label{cons1}
\frac{c_1 \lambda _2 }{\lambda _2 (2 c_2+c_1 t_0 )+2}\geq0.
\end{eqnarray}
In fact, we have already given this constraint under Eq.~(\ref{eq54}), which means that the constraint from the second law of thermodynamics on the coupling parameter $\lambda_2$ is not more stringent than the constraint from the { solution~(\ref{eq54}) itself}.

Moreover, for a closed thermodynamic system, the thermodynamic state tends, inevitably, toward macroscopic equilibrium. In other words, all the macroscopic variables of a closed thermodynamic system always spontaneously approach to constants as time goes on, which implies that there is another constraint on the total entropy $S_{in}(t)$, i.e., $\ddot S_{in}(t\rightarrow \infty)\rightarrow0$.\footnote{Here, we do not take the constraint $\ddot S_{in}<0$, which has been used in most of similar works~\cite{Callen:xx,Mimoso:2013zhp,Harko:2014pqa,Harko:2015pma}. Since the constraint $\ddot S_{in}<0$ for a closed thermodynamic system is only imposed on the last stage of the evolution, we  hold the opinion that $\ddot S_{in}>0$ for small value of $t$ is reasonable (our constraint is more relaxed). If the scale factor corresponds to early or middle universe, such a constraint $\ddot S_{in}<0$ (also $\ddot S_{in}(t\rightarrow \infty)\rightarrow0$) may be needless.} In this cosmological model ($a(t)\propto t^{3/4}$), this requirement for the total entropy is
\begin{eqnarray}
\ddot S_{in}(t\rightarrow \infty)=\lim_{t \to \infty}S_{in}(t_0) \frac{6 \lambda _2^2 c_1^2 [\lambda _2 (2 c_2+c_1 t )+2]}{[\lambda _2 (2 c_2+c_1 t_0)+2]^3}\rightarrow0.
\end{eqnarray}
Obviously, to meet the requirement, we need $\lambda _2=0$ or $c_1=0$. { Both conditions} result in $\Gamma_\phi=0$ and the minimal coupling between the scalar field and curvature. However, if the scale factor ($a(t)\propto t^{3/4}$) does not correspond to the late of the universe, this requirement can be discarded. As we know, it is a generally received opinion that the late universe will be dominated by vacuum energy and the scale factor will grow exponentially over time. For this solution of the scale factor, we can ignore the requirement $\ddot S_{in}(t\rightarrow \infty)\rightarrow0$ and deem that the only constraint on the coupling parameter $\lambda_2$ is (\ref{cons1}), which, unfortunately,
does not provide more { constraints} than that from { the solution~(\ref{eq54}) itself.}

\subsection{$\lambda_1\neq0$ and $\lambda_2=0$}
\label{B}
\subsubsection{$a(t)\propto e^{h t}$}
For $\lambda_2=0$ and $\lambda_1\neq0$, if the scale factor $a(t)$ takes { the} exponential form $a(t)\propto e^{h t}$ and $h$ satisfies $h^2=\frac{1}{3\lambda_1}$, there exists { a trivial analytical solution}
\begin{subequations}
\begin{eqnarray}
a(t)&\propto& e^{\sqrt{\frac{1}{3\lambda_1}}t},\\
V(\phi)&=&-\frac{3}{\lambda_1},\\
\phi(t)&=&c_1e^{\sqrt{\frac{1}{\lambda_1}}t}+c_2,
\end{eqnarray}
\end{subequations}
where $c_1$ and $c_2$ are arbitrary constants and $\lambda_1>0$. Substituting the solutions back into Eq.~(\ref{hornconv}), the energy conservation condition of the scalar field is always tenable, which provides no additional constraint on the coupling parameter $\lambda_1$.

\subsubsection{$a(t)\propto t$}
We consider another case that the scale factor is proportional to $t^n$. In this case, we just find { an analytical solution:}
\begin{subequations}\label{solution_a_t}
\begin{eqnarray}
a(t)&\propto& t,\\
V(\phi(t))&=&\frac{-6\lambda_1+e^{\frac{t^2}{2\lambda_1}}(t^2+3\lambda_1)
\Big(\Gamma [0,\frac{t^2}{2 \lambda_1}]-c_1\Big)}{2\lambda_1t^2},\\
\phi(t)&=&\pm\frac1{\lambda_1}\int \sqrt{e^{\frac{t^2}{2\lambda_1}}\lambda_1\Big(c_1-\Gamma [0,\frac{t^2}{2 \lambda_1}]\Big)}  dt,
\end{eqnarray}
\end{subequations}
where $c_1$ is a constant, and $\Gamma [0,\frac{t^2}{2 \lambda_1}]$ is an incomplete Gamma function. To guarantee that $\Gamma [0,\frac{t^2}{2 \lambda_1}]$ is a real number, we set the parameter $\lambda_1$ to be positive. Note that $\Gamma [0,\frac{t^2}{2 \lambda_1}]$ is always positive and decreases monotonically from $t=0$ ($\Gamma [0,0]\rightarrow +\infty$) to $t=\infty$ ($\Gamma [0,+\infty]\rightarrow 0$). When $t$ is small enough, the solution of the scalar field will be an imaginary number. {Therefore, the} value of $c_1$ could be treated as a truncation parameter, which determines the minimum value of {the time} $t$. When $t$ goes over the bound, the solution of the scalar field is meaningless.

Taking the { solution (\ref{solution_a_t})} into Eq.~(\ref{productionrate3}), we have
\begin{eqnarray}
\Gamma_{\phi}&=&\frac{3 \lambda_1 e^{-\frac{t^2}{2 \lambda_1}}}{t^3 \big(c_1-\Gamma [0,\frac{t^2}{2 \lambda _1}] \big)}+\frac{3 \lambda_1}{t^3}+\frac{3}{2 t}.\label{productionrate33}
\end{eqnarray}
If $\lambda_1>0$ and $t$ is lager than its minimum value, $\Gamma_{\phi}$ is always positive. { The total entropy $S_{in}$ is calculated as}
\begin{eqnarray}
S_{in}(t)=S_{in}(0) \exp\Bigg[\int_{t_0}^{t}\Bigg( \frac{3 \lambda_1 e^{-\frac{t^2}{2 \lambda_1}}}{t^3 \big(c_1-\Gamma [0,\frac{t^2}{2 \lambda_1}] \big)}+\frac{3 \lambda_1}{t^3}+\frac{3}{2 t}\Bigg) dt\Bigg],
\end{eqnarray}
which should satisfy
\begin{eqnarray}\label{cons2}
\frac{dS_{in}(t)}{dt}&=&S_{in}(t)\Bigg[\frac{3 \lambda_1 e^{-\frac{t^2}{2 \lambda_1}}}{t^3 \big(c_1-\Gamma [0,\frac{t^2}{2 \lambda_1}] \big)}+\frac{3 \lambda_1}{t^3}+\frac{3}{2 t}\Bigg]\nonumber\\
&=& S_{in}(t)\Gamma_{\phi}\geq0.
\end{eqnarray}
Because $S_{in}(t)$ is always larger than zero, the above constraint condition is equivalent to $\Gamma_{\phi}\geq0$. From Eq.~(\ref{cons2}) we can draw a conclusion that as long as the total entropy and particle production rate satisfy relation (\ref{ds1}), the constraint on the system from the second law of thermodynamics is equivalent to requiring the particle production rate to be greater than zero, i.e., particles can only be generated.

If the system achieves stability in the form of $a(t)=t$, the second derivative of the entropy $S_{in}(t)$ with respect to the time $t$ needs to satisfy
\begin{eqnarray}
\ddot S_{in}(t\rightarrow \infty)&=&\lim_{t \to \infty}S_{in}(t)(\Gamma_{\phi}^2+\dot{\Gamma}_{\phi})\nonumber\\
&=&\lim_{t \to \infty}S_{in}(t_0)\exp\Big[\int_{t_0}^{t}\Gamma_{\phi}dt\Big](\Gamma_{\phi}^2+\dot{\Gamma}_{\phi})
\rightarrow0.
\end{eqnarray}
We briefly explain that the above requirement is always satisfied for any given $c_1$ and $\lambda_1$. { First of all, $S_{in}(t_0)$ can be ignored since it is just a constant. Secondly,} through a simple calculation we can find that when $t\rightarrow \infty$, the numerator and denominator of $\Gamma_{\phi}^2+\dot{\Gamma}_{\phi}$ are all infinite quantities. But according to L'Hospital law we could tell that $\Gamma_{\phi}^2+\dot{\Gamma}_{\phi}\rightarrow0$ as $t\rightarrow\infty$. At last, the slightly more complex problem is to prove that $(\Gamma_{\phi}^2+\dot{\Gamma}_{\phi})\exp\Big[\int_{t_0}^{\infty}\Gamma_{\phi}dt\Big]$ tends to 0 when $t\rightarrow\infty$. Reviewing Eq.~(\ref{productionrate33}), we can use a special function to replace $\Gamma [0,\frac{t^2}{2 \lambda _1}]$ and then prove it. Since $\Gamma_{\phi}$ is always positive, we have
\begin{eqnarray}
\exp\Big[\int_{t_0}^{\infty}\Gamma_{\phi}dt\Big]&=&\exp\Big[\int_{t_0}^{\infty}\Bigg(\frac{3 \lambda_1 e^{-\frac{t^2}{2 \lambda_1}}}{t^3 \big(c_1-\Gamma [0,\frac{t^2}{2 \lambda _1}] \big)}+\frac{3 \lambda_1}{t^3}+\frac{3}{2 t}\Bigg)dt\Big]\nonumber\\
&<&\exp\Big[\int_{t_0}^{t_1}\frac{3 \lambda_1 e^{-\frac{t^2}{2 \lambda_1}}}{t^3 \big(c_1-\Gamma [0,\frac{t^2}{2 \lambda _1}] \big)}dt
+\int_{t_1}^{\infty}\frac{3 \lambda_1 e^{-\frac{t^2}{2 \lambda_1}}}{t^3 \big(c_1-\frac{1}{t^2} \big)}dt\nonumber\\
&+&\int_{t_0}^{\infty}\Bigg(\frac{3 \lambda_1}{t^3}+\frac{3}{2 t}\Bigg)dt\Big].
\end{eqnarray}
For any given $\lambda_1$ there always exists a finite $t_1$, which guarantees $\Gamma [0,\frac{t^2}{2 \lambda _1}]<\frac{1}{t^2}<c_1$ when $t>t_1$. { It can be shown} that the above integral result is equivalent to $C(t_1,t_0) \exp[\int_{t_0}^{\infty}\frac{3}{2 t}dt]$, where $C(t_1,t_0)$ is a constant. Although the integral $\exp[\int_{t_0}^{\infty}\frac{3}{2 t}dt]$ is divergent, using L'Hospital law again one can prove { that} $(\Gamma_{\phi}^2+\dot{\Gamma}_{\phi})\exp\Big[\int_{t_0}^{\infty}\Gamma_{\phi}dt\Big]$ is not divergent and { goes to zero} as $t\rightarrow\infty$.

Comparing the former case ($\lambda_1=0$, $\lambda_2\neq0$ and $a(t)\propto t^{3/4}$) with this case, we find that the second law of thermodynamics is always satisfied in these two cases. The requirement for the stability of the system either results in $\Gamma_{\phi}=0$ ($\lambda_1=0$, $\lambda_2\neq0$ and $a(t)\propto t^{3/4}$) or cannot provide new constraints ($\lambda_1\neq0$, $\lambda_2=0$ and $a(t)\propto t$), which are not the results {we expect.}

\subsection{$\lambda_1\neq0$ and $\lambda_2\neq0$}
\label{C}
\subsubsection{$a(t)\propto e^{h t}$}
When $\lambda_2\neq0$, $\lambda_1\neq0$ and $a(t)\propto e^{ht}$, the {solution of Eq.~(\ref{motion4}) is} given as
\begin{subequations}
\begin{eqnarray}
a(t)&\propto& e^{\frac{2}{3}\sqrt{\frac{1}{\lambda_1}}t},\\
\phi(t)&=&b t,\\
V(\phi(t))&=&\frac{b^2}2-\frac{12+2b\sqrt{\lambda_1}}{9\lambda_1}\phi=
\frac{b^2}2-\frac{{12b}+2b^2\sqrt{\lambda_1}}{9\lambda_1}t,\\
F(\phi(t))&=&\frac{b\sqrt{\lambda_1}}{6\lambda_2}\phi=\frac{b^2\sqrt{\lambda_1}}{6\lambda_2}t,
\end{eqnarray}\label{solution1}
\end{subequations}
where $\lambda_1>0$. The scalar particle production rate is
\begin{eqnarray}
\Gamma_{\phi}&=&\frac{16}{9} \sqrt{\frac{1}{\lambda_1}}>0.\label{productionrate44}
\end{eqnarray}
Note that $\Gamma_{\phi}$ is a constant and it is only related to {the} coupling parameter $\lambda_1$.
For this solution the total entropy of the system is given by
\begin{eqnarray}
S_{in}(t)=S_{in}(t_0) \exp\left(\frac{16}{9} \sqrt{\frac{1}{\lambda_1}}t\right)>0,
\end{eqnarray}
which leads to {
\begin{eqnarray}
\ddot S_{in}(t\rightarrow \infty)
 &=&\lim_{t \to \infty}S_{in}(t_0)\frac{256}{81\lambda_1} \exp\left(\frac{16}{9} \sqrt{\frac{1}{\lambda_1}}t\right)\rightarrow\infty.
\end{eqnarray} }
The result is usually unacceptable. In summary, the thermodynamic requirements do not provide extra constraints for the coupling parameters.

\subsubsection{$a(t)\propto t^{3/4}$}
For the stage that the expansion of the universe is dominated by the dust or radiation, we set $a(t)\propto t^n$. We just find {the following analytical solution:}
\begin{subequations}
\begin{eqnarray}
a(t)&\propto& t^{\frac{3}{4}},\\
\phi(t)&=&p t^{\frac{19}{8}}+q,\\
V(\phi(t))&=&\frac{9 \lambda_2}{32}t^{-\frac{9}{8}} \left[(q-\sqrt{47} d) \sin \bigg(\frac{\sqrt{47}}{8} \log (t)\bigg)+(d+\sqrt{47} q) \cos \bigg(\frac{\sqrt{47}}{8} \log (t)\bigg)\right]\nonumber\\
&+&\frac{207}{544} \lambda _1 p^2 t^{3/4}-\frac{15}{14} p^2 t^{11/4},\\
F(\phi(t))&=&t^{\frac{7}{8}}\left[d \cos \bigg(\frac{\sqrt{47} }{8} \log (t)\bigg)+q \sin \bigg(\frac{\sqrt{47}}{8} \log (t)\bigg)\right]-\frac{1}{\lambda_2}+\frac{27 \lambda _1 p^2 t^{11/4}}{68 \lambda _2}\nonumber\\
&-&\frac{4 p^2 t^{19/4}}{63 \lambda _2},
\end{eqnarray}\label{solution2}
\end{subequations}
where $p$, $q$ and $d$ are constants. For simplicity, we only consider the case with $q=d=0$, which leads $\Gamma_{\phi}$ to be independent of $\lambda_2$, but it is enough to illustrate the problems. The scalar particle production rate $\Gamma_{\phi}$ and the total entropy $S_{in}$ of the system are quite simple:
\begin{eqnarray}
\Gamma_{\phi}&=&\frac{19}{28 t}+\frac{621 \lambda _1}{2176 t^3},\label{gamma3}\\
S_{in}(t)&=&S_{in}(t_0) t^{\frac{19}{28}} e^{-\frac{621 \lambda _1}{4352 t^2}}.
\end{eqnarray}
The second law of thermodynamics requires $\Gamma_{\phi}>0$, which means $\lambda_1>0$. Since this requirement for $\lambda_1$ is not necessary in the original equation and solution, we do obtain a new restraint on { the} coupling parameters by considering thermodynamic properties of the system.

The second derivative of $S_{in}$ with respect to $t$ is
\begin{eqnarray}
\ddot S_{in}(t)=- S_{in}(t_0)\frac{9 e^{-\frac{621 \lambda _1}{4352 t^2}} \Big[5622784 t^4+11109 \lambda _1 (1088 t^2-189 \lambda _1)\Big]}{232013824 t^{149/28}}.
\end{eqnarray}
It can be shown that when the time $t$ tends to infinity, $\ddot S_{in}(t)$ will approach to zero.
\subsubsection{$a(t)\propto t^{n}$ $(n\neq3/4)$}
As for other cases ($a(t)\propto t^{n}$ and $n\neq3/4$), it is very difficult to solve Eq.~(\ref{motion4}) analytically. To study the scale factor taking the forms of $a(t)\propto t^{\frac{1}{2}}$ and $a(t)\propto t^{\frac{2}{3}}$, there are two possible ways: numerical or approximate solutions. Here, we choose the later way.

When $a(t)\propto t^{\frac{1}{2}}$, we can simplify Eq.~(\ref{motion4}) before taking the approximation. We find that if $\phi(t)$ and $V(\phi)$ satisfy, respectively, the following relations:
\begin{subequations}
\begin{eqnarray}
\dot{\phi}(t)&=&\sqrt{\frac{-2 \lambda_2 t \big[2 t \ddot{f}(t)+\dot{f}(t)\big]-4 \lambda_2 f(t)-4}{4 t^2-3 \lambda_1}},\\
V(\phi)&=&-\frac{6 + 6 \lambda_2 \big[f(t)-2t\dot{f}(t)\big]+(4t^2+9\lambda_1)\dot{f}(t)^2}{8 t^2},
\end{eqnarray}
\end{subequations}
{then} the three equations in (\ref{motion4}) can be reduced to a higher-order equation of $f(t)$:
\begin{eqnarray}\label{eq533}
16 \lambda _1 \lambda _2 t^5 f^{(3)}(t)-12 \lambda _1^2 \lambda _2 t^3 f^{(3)}(t)-32 \lambda _1 \lambda _2 t^4 \ddot{f}(t)-32 \lambda _2 t^5 \dot{f}(t)+36 \lambda _1 \lambda _2 t^3 \dot{f}(t)\nonumber\\
-21 \lambda _1^2 \lambda _2 t \dot{f}(t)-72 \lambda _1 \lambda _2 t^2 f(t)+30 \lambda _1^2 \lambda _2 f(t)+30 \lambda _1^2-72 \lambda _1 t^2=0.
\end{eqnarray}
If this higher-order equation has an analytical solution, the whole equations (\ref{motion4}) will have analytical solutions. Unfortunately, it is almost impossible to find an analytical solution for this equation. The approximation we take depends on the values of the parameters $\lambda_1$ and $\lambda_2$. First, we suppose that the coupling parameters $\lambda_1$ and $\lambda_2$ are all small quantities, so we may ignore {the} second and higher order small quantities in Eq.~(\ref{eq533}), which results in the following approximation equation:
\begin{eqnarray}
-32 \lambda _2 t^5 \dot{f}(t)-72 \lambda _1 t^2=0.
\end{eqnarray}
Our approximate solutions are given as
\begin{subequations}
\begin{eqnarray}
a(t)&\propto& t^{\frac{1}{2}},\\
\dot{\phi}(t)&=&\sqrt{\frac{-4 t^2 \left(c_1 \lambda _2+1\right)-27 \lambda _1}{4 t^4-3 \lambda _1 t^2}},\\
V(\phi(t))&=&\frac{-32 t^4 \left(c_1 \lambda _2+1\right)+108 \lambda _1 t^2 \left(2 c_1 \lambda _2+1\right)+1377 \lambda _1^2}{32 t^4 \left(4 t^2-3 \lambda _1\right)},\\
F(\phi(t))&=&\frac{9 \lambda _1}{8 \lambda _2 t^2}+c_1,
\end{eqnarray}
\end{subequations}
where $c_1$ is a constant. To guarantee that the range of time $t$ is as large as possible, we need $c_1 \lambda _2+1<0$. The corresponding particle production rate and entropy are given as
\begin{eqnarray}
\Gamma_{\phi}&=&-\frac{9 \lambda _1 \left(4 t^2-3 \lambda _1\right) \left(4 t^2 \left(c_1 \lambda _2+1\right)+51 \lambda _1\right)}{8 t^3 \left(\lambda _1+4 t^2\right) \left(4 t^2 \left(c_1 \lambda _2+1\right)+27 \lambda _1\right)},\\
S_{in}(t)&=&S_{in}(t_0)\exp \Bigg[\frac{9}{8} \lambda _1 \bigg(\frac{8 \left(c_1 \lambda _2-50\right) \log \left(\lambda _1+4 t^2\right)}{\lambda _1 \left(c_1 \lambda _2-26\right)}-\frac{16 \left(2 c_1 \lambda _2+155\right) \log (t)}{81 \lambda _1}\\
&+&\frac{16 \left(c_1 \lambda _2+1\right) \left(c_1 \lambda _2+10\right) \log \left(4 t^2 \left(c_1 \lambda _2+1\right)+27 \lambda _1\right)}{81 \lambda _1 \left(c_1 \lambda _2-26\right)}-\frac{17}{6 t^2}
\bigg)\Bigg].
\end{eqnarray}
Again we require $\Gamma_{\phi}>0$ and $\ddot{S}_{in}(t\rightarrow\infty)\rightarrow0$, which leads to $\lambda_1<0$.

As for the case of $a(t)\propto t^{\frac{2}{3}}$, by applying the similar approximate treatment, the result is similar, so we do not repeat it here.

\section{Entropy of the apparent horizon}
Now, we briefly discuss the entropy of the apparent horizon of the universe, which can always be given by the radius of the apparent horizon~\cite{Pavon:2012qn,Mimoso:2013zhp,Easson:2010av,Bardeen:1973gs,Hawking:1974sw,Bekenstein:1973ur,
Gibbons:1977mu,Wald:1993nt,Brustein:2007jj}:
\begin{eqnarray}\label{entropyofH}
S_h(t)=\frac{\pi}{G_{eff}H^2},
\end{eqnarray}
where $G_{eff}$ is an effective Newtonian constant. The radius of the apparent horizon is $r_h=H^{-1}$ for the FRW universe with $k=0$~\cite{Bak:1999hd}. One can simply regard $S_h(t)\propto\frac{1}{H^2}$ in modified gravity theories. Apparently, when $a(t)\propto e^{ht}$, $S_h(t)$ is a constant satisfying the thermodynamic constraints $\dot S_h(t)\geq0$ and $\ddot S_h(t\rightarrow\infty)\rightarrow0$. Unfortunately, as we discussed earlier, when $a(t)\propto e^{ht}$, the entropy generated by particle production, usually, does not meet the same requirements $\dot S_{in}(t)\geq0$ and $\ddot S_{in}(t\rightarrow\infty)\rightarrow0$ (at least in our simple Horndeski theory). Therefore, the sum of the entropy $S_{sum}=S_{in}+S_h$ cannot satisfy all thermodynamic constraints. Since the scale factor in the late universe is likely to grow exponentially, it is best not to ignore the second requirement. So the result is that $\Gamma_{\phi}=0$, i.e., all the coupling parameters must be zero.

In the case of $a(t)\propto t^n$, we have $S_h(t)=\pi t^2/n^2$, which guarantees that $\dot S_h(t)\geq0$ but not $\ddot S_h(t\rightarrow\infty)\rightarrow0$. Obviously, the sum of the entropy also cannot satisfy $\ddot S_{sum}(t\rightarrow\infty)\rightarrow0$. As we mentioned earlier that for the case $a(t)\propto t^n$, since it is not the ultimate evolution of the universe, the second requirement can be relaxed.

\section{Conclusion and discussion}
\label{Conclusion}

In the present paper we continued to study particle production {based on the} works~\cite{Harko:2014pqa,Harko:2015pma}. For this purpose, we need to study a theory with non-minimal coupling between matter and space-time. Since Horndeski theory contains couplings of scalar field with space-time and also guarantees that the field equations are second order, we considered a simple Horndeski theory as our main research object.

First, by defining the scalar field energy-momentum tensor with Eq.~(\ref{Tuv}), we obtained the energy conservation relation~(\ref{hornconv}) between the scalar field and geometry. We found that the great advantage of considering a specific action of matter is that we can derive Eq.~(\ref{horndeskiscalar}) directly from the equation of motion of the matter, which could avoid complex calculations when the Einstein's field equations is cumbersome.
Following Harko's method~\cite{Harko:2014pqa,Harko:2015pma}, we applied the non-conservation of the scalar field energy-momentum tensor to an open thermodynamic system and obtained abstract expressions of particle production rate and the corresponding entropy.

Then, we solved the field equations of the simple Horndeski theory analytically in the context of the universe. The solutions were divided into two categories by the form of scale factor: $a(t)\propto e^{ht}$ and $a(t)\propto t^n$. When the coupling parameters $\lambda_1\neq0$ and $\lambda_2\neq0$, we hypothesized that $\lambda_1$ and $\lambda_2$ are all small quantities and got some approximate solutions. With these solutions, we obtained expressions of entropy evolving with time $t$. Since the open cosmological systems we studied are thermodynamic systems, one can naturally put forward two requirements on the entropy of system: $\dot S_{in}(t)\geq0$ and $\ddot S_{in}(t\rightarrow\infty)\rightarrow0$. The first requirement is equivalent to $\Gamma_{\phi}\geq0$, i.e., particles can only be generated. The second requirement $\ddot S_{in}(t\rightarrow\infty)\rightarrow0$ is looser than the constraint $\ddot S_{in}<0$~\cite{Callen:xx,Mimoso:2013zhp,Harko:2014pqa,Harko:2015pma}.

With the help of thermodynamic properties of the system, we hope to give some new constraints on the coupling parameters $\lambda_1$ and $\lambda_2$ in the simple Horndeski theory. Unfortunately, most of our results, for example the case of $\lambda_1=0$, $\lambda_2\neq0$ and $a(t)\propto t^{3/4}$, show that thermodynamic laws cannot provide more constraints than that from the solutions themselves. And the cases which give new constraints on the coupling parameters (for example the case of $\lambda_1\neq0$, $\lambda_2\neq0$ and $a(t)\propto t^{\frac{1}{2}}$) can only limit the sign of them and provide no more specific details. The reason may be that the range of the parameter $t$ is always $(0,\infty)$. If the range of time can be more specific, especially for the case $a(t)\propto t^n$, perhaps the results will be more stringent.

It is worth mentioning that the requirement $\ddot S_{in}(t\rightarrow\infty)\rightarrow0$ (and also $\ddot S_{in}<0$) may be improper when the form of $a(t)$ does not represent the late universe, so this requirement is always negligible unless we really know the final evolution of the universe. But the first requirement $\dot S_{in}(t)\geq0$ is true in any case. Fortunately, sometimes even if we only consider $\dot S_{in}(t)\geq0$, it could give some new constraints on the coupling parameters (see the example of $\lambda_1\neq0$, $\lambda_2\neq0$ and $a(t)\propto t^{\frac{1}{2}}$).

Finally, we briefly discussed the entropy of the apparent horizon. Since Eq.~(\ref{entropyofH}) is too rough, it is not much helpful in our research. In the presence of the coupling of geometry with matter, it is reasonable that the entropy of the apparent horizon is related to the coupling parameters. If so, Eq.~(\ref{entropyofH}) should be recalculated. The new formula of the entropy is promising to provide more constrains and our follow-up work will put this as a priority. What is more, if the background space-time itself has entropy, it is also necessary to consider the entropy of the space-time in the sum of the entropy, which is also worthy of our continued researches.

\section*{Acknowledgements}

This work was supported by the National Natural Science Foundation of China (Grants No. 11522541 and No. 11375075), and the Fundamental Research Funds for the
Central Universities (Grant No. lzujbky-2017-it68 and No. lzujbky-2015-jl01).
H. Yu was supported by the scholarship granted by the Chinese Scholarship Council (CSC).

\end{document}